\documentclass[twocolumn]{article}
\usepackage[utf8]{inputenc}
\usepackage[dvipdfmx]{graphicx}
\usepackage{siunitx}
\usepackage{authblk}
\usepackage{hyperref}
\usepackage{amsmath}
\usepackage{bm}

\usepackage{upgreek}
\usepackage{gensymb}

\hypersetup{
    colorlinks=true,
    linkcolor=blue,
    filecolor=magenta,      
    urlcolor=cyan,
    }

\usepackage[margin=2.5cm]{geometry}

\title{\vspace{-15mm} Contactless pressure measurement of an underwater shock wave in a microtube using a high-resolution background-oriented schlieren technique}

\author[1]{\small{Shota Yamamoto}}
\author[1]{Takaaki Shimazaki}
\author[2]{Andr\'es Franco-G\'omez}
\author[1]{Sayaka Ichihara}
\author[1]{Jingzu Yee}
\author[1,3]{Yoshiyuki Tagawa}

\affil[1]{Department of Mechanical Systems Engineering, Tokyo University of Agriculture and Technology, Koganei Campus 6-204, 2-24-16 Nakacho, Koganei, Tokyo, Japan
}
\affil[2]{Department of Chemical Engineering, Biotechnology and Materials, FCFM, University of Chile, Santiago, Chile}

\affil[3]{Institute of Global Innovation Research, Tokyo University of Agriculture and Technology, Koganei Campus 6-204, 2-24-16 Nakacho, Koganei, Tokyo, Japan
}

\date{}

\begin{document}

\twocolumn[
\maketitle
\vspace{-10mm}
\paragraph{Abstract} A high-resolution background-oriented schlieren (BOS) technique, which utilizes a high-resolution camera and a microdot background pattern, is proposed and used to measure the pressure field of an underwater shock wave in a microtube.
The propagation of the shock wave subsequently reaches a concave water-air interface set in the microtube resulting in the ejection of a focused microjet.
This high spatial-resolution BOS technique can measure the pressure field of a shock front with a width as narrow as the order of only $10^1$~$\upmu$m with a peak pressure as large as almost $3$ MPa, which is significantly narrower and larger, respectively, than a previous study \cite{Hayasaka2016}.
This significant breakthrough has enabled the simultaneous measurement of the pressure impulse of the shock front and the velocity of the microjet tip.
As a result, we have experimentally observed the linear relation between the velocity of the microjet tip and the pressure impulse of the shock front for the cases without secondary cavitation in the liquid bulk. 
Such relation was theorectically/numerically predicted by Peters \cite{Peters2013}.
This study demonstrated the capability of the proposed high-resolution BOS technique as a microscale contactless pressure measurement tool for underwater shock waves and potentially other micro- and nanofluids.
\vspace{20pt}
]

\section{Introduction}
Underwater shock waves generated in a microtube are essential for the rapid transport of liquids in microfluidic devices.
Typical examples include the generation of high-speed microjets for needle-free injection systems \cite{Menezes2009}.
As predicted theoretically/numerically, the microjet velocity is strongly influenced by the pressure field in microfluidic devices (\cite{Peters2013, Tagawa2012}).
However, the use of contact measurement devices, such as needle-type hydrophones, causes distortion to the minute-scaled flow.
To measure the pressure in microfluidic devices, we focused on the background-oriented schlieren (BOS) technique \cite{Venkatakrishnan2004}, which is a contactless density measurement technique with a simple experimental setup \cite{hayasaka2019mobile}. 
The BOS technique has been applied to measure the density field of a shock wave (\cite{meier2002computerized,Yamamoto2015}).
The first successful pressure field measurement using the BOS technique was performed by \cite{Hayasaka2016}.
However, to measure the pressure field in microfluidic devices, the spatial resolution must be significantly higher than $10$ $\upmu$m/pixel (\cite{Hayasaka2016}) to accurately measure the distortion of the displacement field.
In this letter, we apply the noninvasive BOS technique by increasing the spatial resolution to $0.68$ $\upmu$m/pixel, which is 15 times higher than that of \cite{Hayasaka2016}.
We then focus on an analysis between the pressure impulse and microjet velocity generated by the corresponding underwater shock waves.
Note that the pressure impulse of the underwater shock wave is obtained by the integration of the pressure field in the shock front vicinity with respect to time.

\section{Experimental methods}\label{Methods}
\subsection{Experimental setup}
\begin{figure}
\begin{center}
\includegraphics[width=1.0\columnwidth]{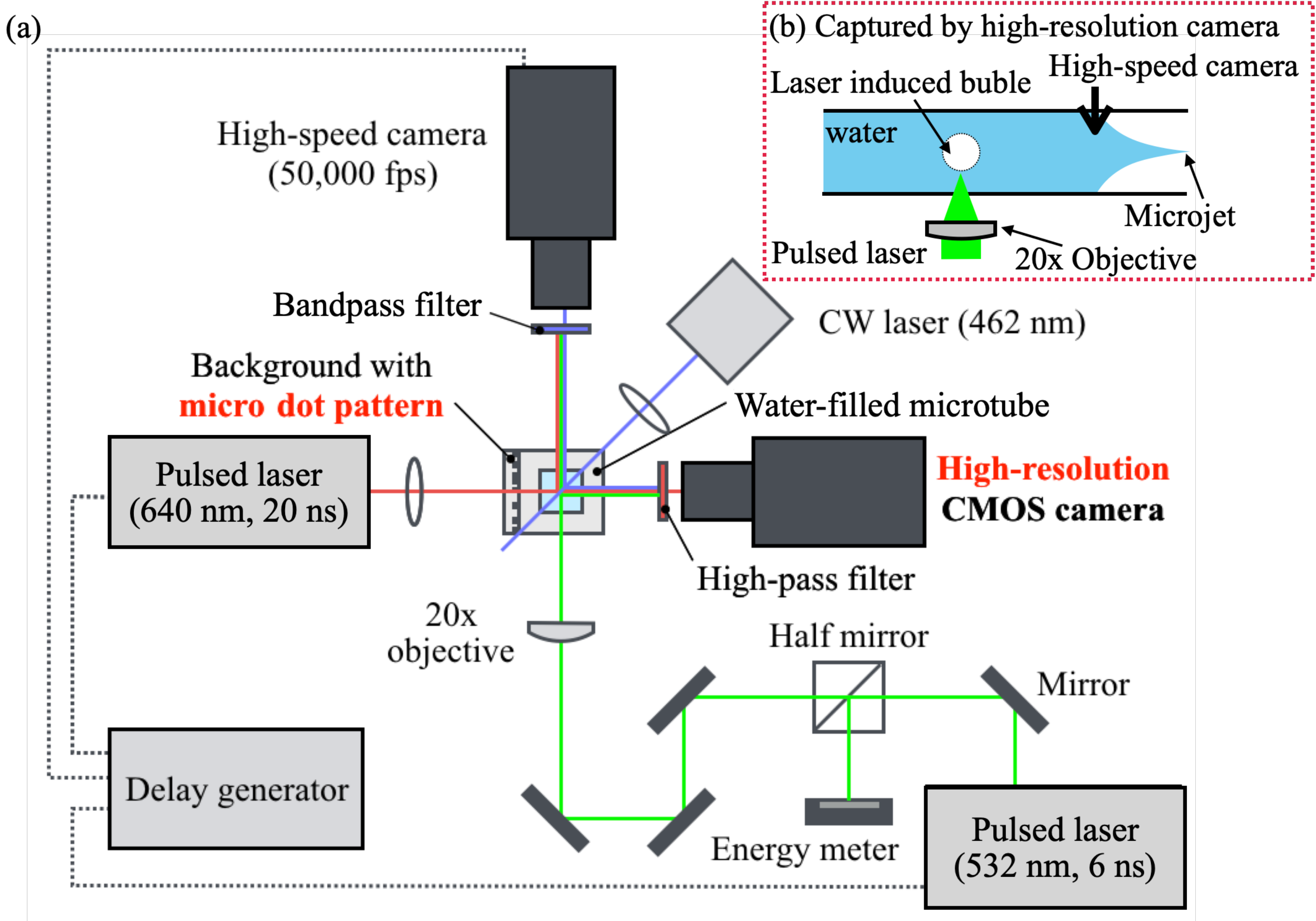}
\caption{(a) Schematic diagram of the experimental setup. A laser induced
  cavitation bubble is generated in a microtube filled with water which provokes the ejection of a microjet from the fluid interface. 
  A pulsed laser of $640$ nm and a high resolution (CMOS) camera are used to capture the displacement field in the water using a microdot background, while a high speed camera captures the jet ejection.
  (b) Schematic diagram of the microjet generation in the microtube.}
\vspace{-5mm}
\label{fig:setup}
\end{center}
\end{figure}

A transparent square capillary tube of $500$ $\upmu$m inner width, $250$ $\upmu$m wall thickness, and $50$ mm length is partially filled with ultra-pure water where a pulsed laser (Nd: YAG laser Nano S PIV, Litron $532$ nm, $6$ ns pulse) is focused through an objective lens (SLMPLN 20X, Olympus) to generate an underwater shock wave (see Fig. \ref{fig:setup}).
In our experiments, the energy of a single pulse is varied from $0.36$ mJ
to $0.97$ mJ.
The microdot background pattern (MEMS 8$\times$8 $\upmu$m dot size) is attached to the wall of the capillary tube. %, see Fig. \ref{fig:background}
High resolution images are captured using a CMOS camera (EOS Kiss X5, Canon, photographing resolution: $4,000-6,000$ pixels) while being illuminated by an expanded laser beam of $640$ nm (SI-LUX 640, Specialized Imaging Ltd.).
The spatial resolution is set at 0.68 $\upmu$m/pixel, which has enabled us to measure the pressure field of a shock front of a width as narrow as the order of $10^1$~$\upmu$m with a peak pressure as large as approximately $3$ MPa, which is significantly narrower and larger, respectively, than the previous studies (\cite{Hayasaka2016} and \cite{Yamamoto2015}, i.e., $1$ MPa.)

The microjet is captured with a high-speed camera at $50,000$ fps (FASTCAM SA-X2, Photron) and illuminated using an expanded laser of $462$ nm (CW Laser, IL-106B, HARDsoft).
The shock wave laser, the high-resolution camera, the high-speed camera, and the light sources are triggered using a delay generator (Model 575, BNC).
To avoid cross illumination between the light sources, we install a high-pass filter on the high-resolution camera and a bandpass filter on the high-speed camera.
An energy meter (EnergyMax-RS J0MB-HE, Coherent, USA, measurable range: 12 $\upmu$J-Hz) measuring the irradiation energy of the laser pulse using a half mirror (OptoSigma, transmittance: $50$\% up to $20$ mJ) is assumed to be proportional to the energy absorbed by the irradiated water volume.
% \vspace{-4mm}

\subsection{BOS analysis}
% \vspace{-4mm}
%------------ figure ------------------------------
%----------------------------------------------------

\begin{figure}
\begin{center}
\includegraphics[width=1.0\columnwidth]{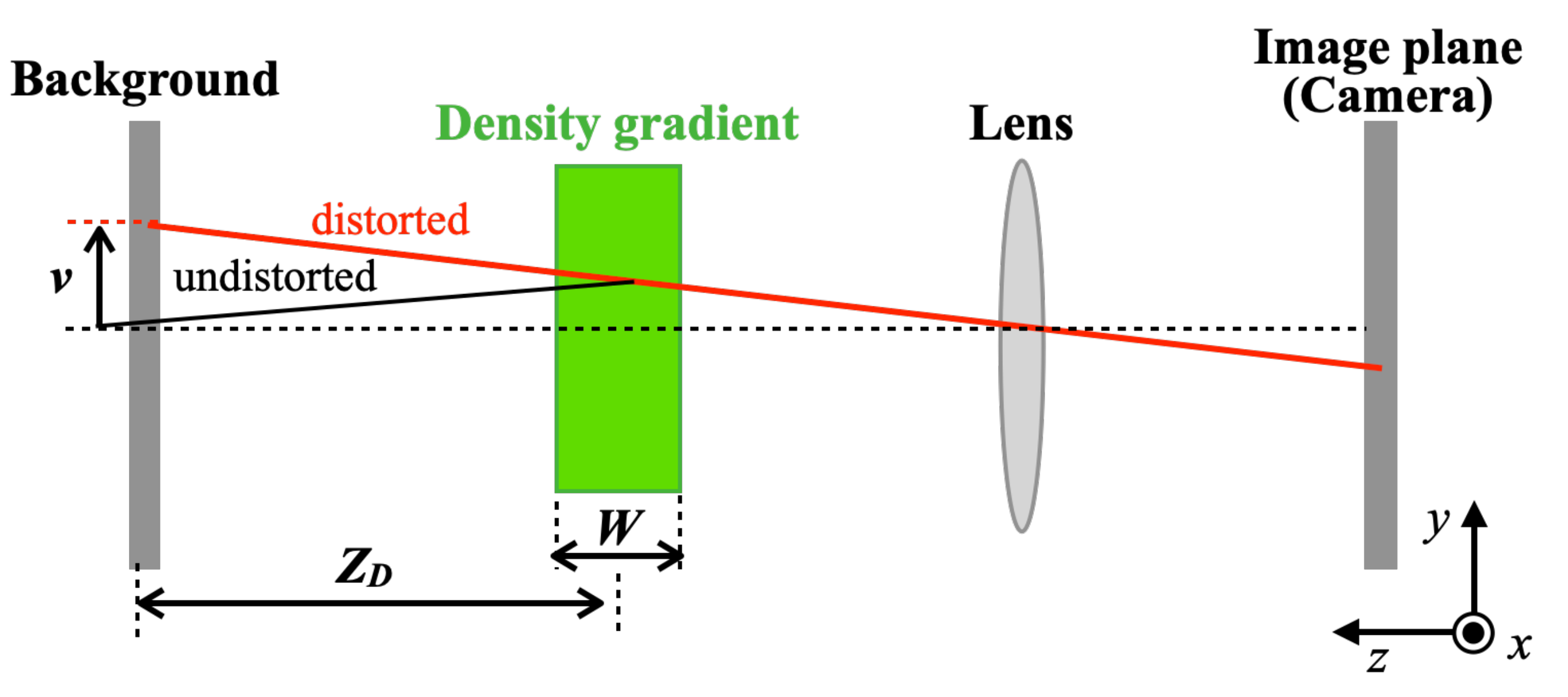} 
%height=4.3cm]{Principle_BOS_update2.pdf}
\caption{The micro background and the image plane are aligned along the optical axis, i.e., the $z$-axis. With a distorted density field of depth $W$ and a distance $Z_D$, a displacement $v$ is observed in the image plane (red dashed path). The solid black line indicates an undistorted density field.}
% \vspace{-6mm}
\label{fig:principle}
\end{center}
\end{figure}

As illustrated in Figure \ref{fig:principle}, the BOS technique quantifies a density-gradient field of a target by comparing the background image distorted by the density gradient of the target and an undistorted background image.
\cite{Venkatakrishnan2004} showed that the displacement field of the distortion is proportional to the gradient of the refractive index.
Therefore, it can be converted into the fluid density and then into the pressure distribution of the fluid.
For this, the theory of the BOS technique was previously developed in detail by \cite{Hayasaka2016}.

The measurement accuracy of the underwater shock wave pressure is highly dependent on the spatial resolution of the BOS technique.
The change in the local displacement as a function of position can be written as $\sim\langle\partial v / \partial y\rangle \Delta y$, where $\Delta y$ is the spatial resolution of the image. 
$\langle\partial v / \partial y\rangle \Delta y$ should be smaller than a certain threshold $\delta v_{th}$.
By comparing BOS and hydrophone measurements \cite{Yamamoto2015,shimazaki2022}, we estimate that $\delta v_{th}\leq0.4$.
Therefore, to increase displacement field gradients (larger pressures), a smaller spatial resolution, such as $\Delta y\sim \delta v_{th}/\langle\partial v / \partial y\rangle$ is required.

\begin{figure}[!ht]
\begin{center}
\includegraphics[width=1.0\columnwidth]{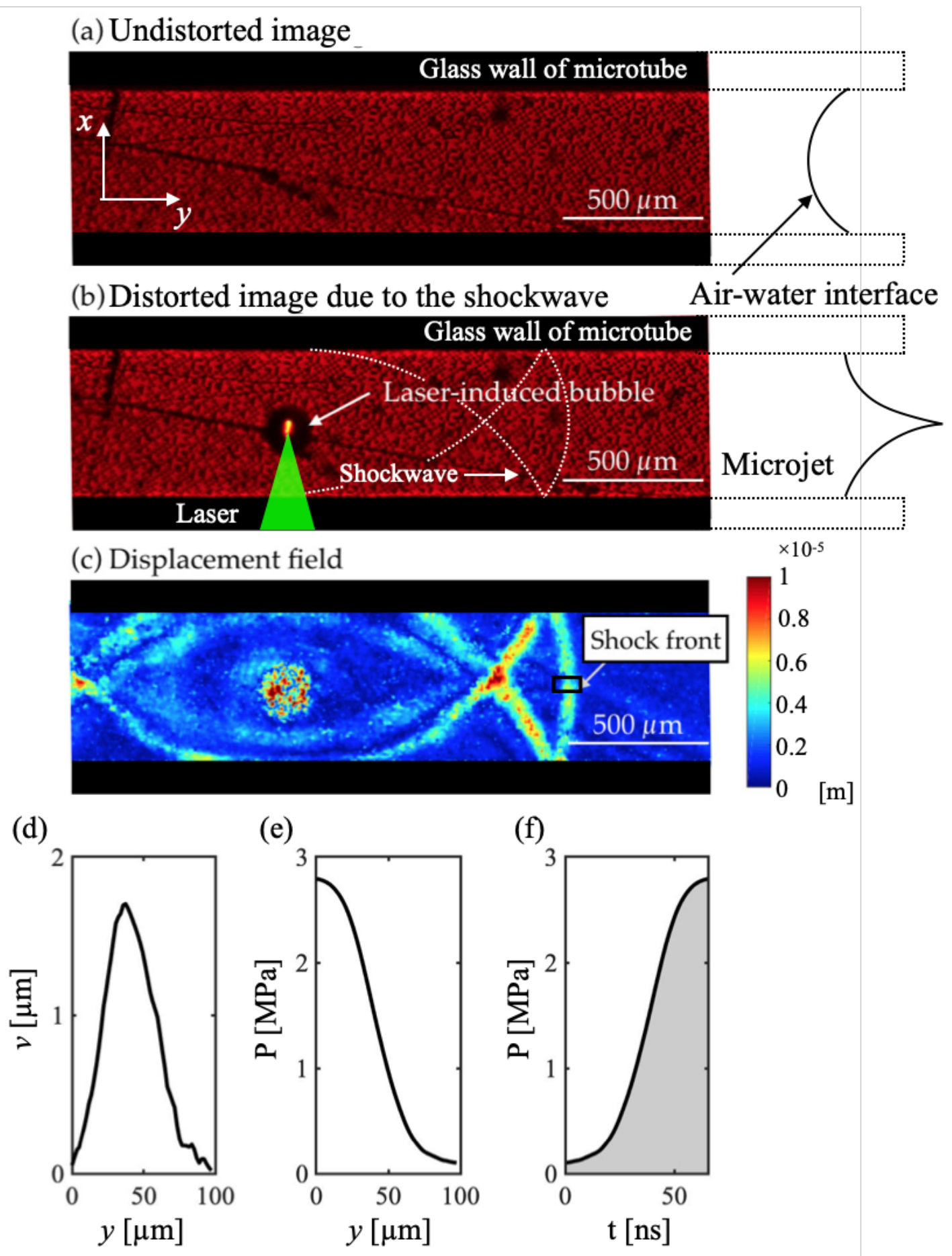}
\caption{(a) Undistorted background image and (b) background image distorted by the underwater shock wave captured by the high-resolution camera. (c) Displacement field obtained from images (a) and (b) using PIV analysis. (d) Averaged profile of displacement field obtained from (c) in the shock front (indicated by the square region with black outline). (e) Pressure profile as a function of distance obtained from the averaged displacement field, and (f) pressure as a function of time obtained from (e), where the area under the curve corresponds to the pressure impulse.}
% \vspace{-6mm}
\label{fig:bos}
\end{center}
\end{figure}

Typical high-resolution BOS images of the backgrounds without the shockwave (undistorted) and with the shock wave (distorted) obtained with our system are presented in Figures~\ref{fig:bos}(a) and (b), respectively.
After the laser pulse was triggered ($t = 0$ $\upmu$s), the image of each shock wave was taken at an interval of $t= 0.5$ $\upmu$s.
The displacement field shown in Fig. \ref{fig:bos}(c) is obtained by accurately registering (digital matching) the distorted and undistorted background images and computing the PIV \nobreak cross-correlation (PIVlab, MATLAB2018), where the overall shape of the micro shock wave is captured.

In this letter, we are interested in measuring the localized pressure at the shock front (indicated by the square region with black outline in Figure \ref{fig:bos}(c)).
The displacement field of the localized region of interest is averaged along the $x$-axis for noise reduction and the profile is shown in Figure \ref{fig:bos}(d).
Along the $z$-axis, the profile is assumed to be constant to simplify the calculation.
The density gradient is obtained by substituting the displacements $v$ ($\displaystyle \langle v\rangle_x \propto W \langle \partial \rho/\partial y \rangle_x$) from Figure \ref{fig:bos}(d) into the Gladstone-Dale equation (\cite{Yamamoto2015}). 
We then obtained the fluid density by integrating the density gradient with respect to the undistorted area (hydrostatic density $\rho_0=998$ kg/m$^3$),
($\rho-\rho_0) \propto Z_D W \int \langle v\rangle_x\,dy$, 
where shock waves do not propagate (right of the shock front shown in Fig. \ref{fig:bos}(c)) (see \cite{Hayasaka2016}).
From the density, the pressure value is calculated through the Tait equation, $\frac{p+T}{p_0+T}=\Bigl(\frac{\rho}{\rho_0}\Bigl)^\beta$,
where $p_0$ is the hydrostatic pressure, and $T$ and $\beta$ is a constant with values of 314 MPa and 7, respectively.
The spatial distribution of the pressure at the shockwave is shown in Figures \ref{fig:bos}(e).
It is transformed into the temporal distribution shown in Figures \ref{fig:bos}(f) by assuming that the (unchanging) shock front is translating along the $y$-axis at the propagation speed of sound in water (1482 m/s at 20$\degree$C).
The pressure impulse of the shock wave is then obtained by integrating the temporal distribution, as illustrated in Figure \ref{fig:bos}(f).

The numerical simulations performed by \cite{Peters2013} have shown that the pressure impulse is proportional to the microjet velocity.
Note that this simulation did not consider the effect of secondary cavitation in a liquid bulk, which might increase the jet velocity (\cite{Kiyama2016, Hayasaka2017}).
% \vspace{-7mm}

\section{Results and discussion}\label{Results}

The shock wave in the microtube is successfully captured by the high-resolution BOS in all experiments, and an example of the results is shown in Figure \ref{fig:bos}.
As shown in Fig. \ref{fig:bos}(e), the measured shock front is as narrow as approximately 90 $\upmu$m.
The peak pressure of the shock front is almost as large as 3 MPa, which is approximately 3 times larger than \cite{Yamamoto2015}.

\begin{figure}[!ht]
\begin{center}
\includegraphics[width=1.0\columnwidth]{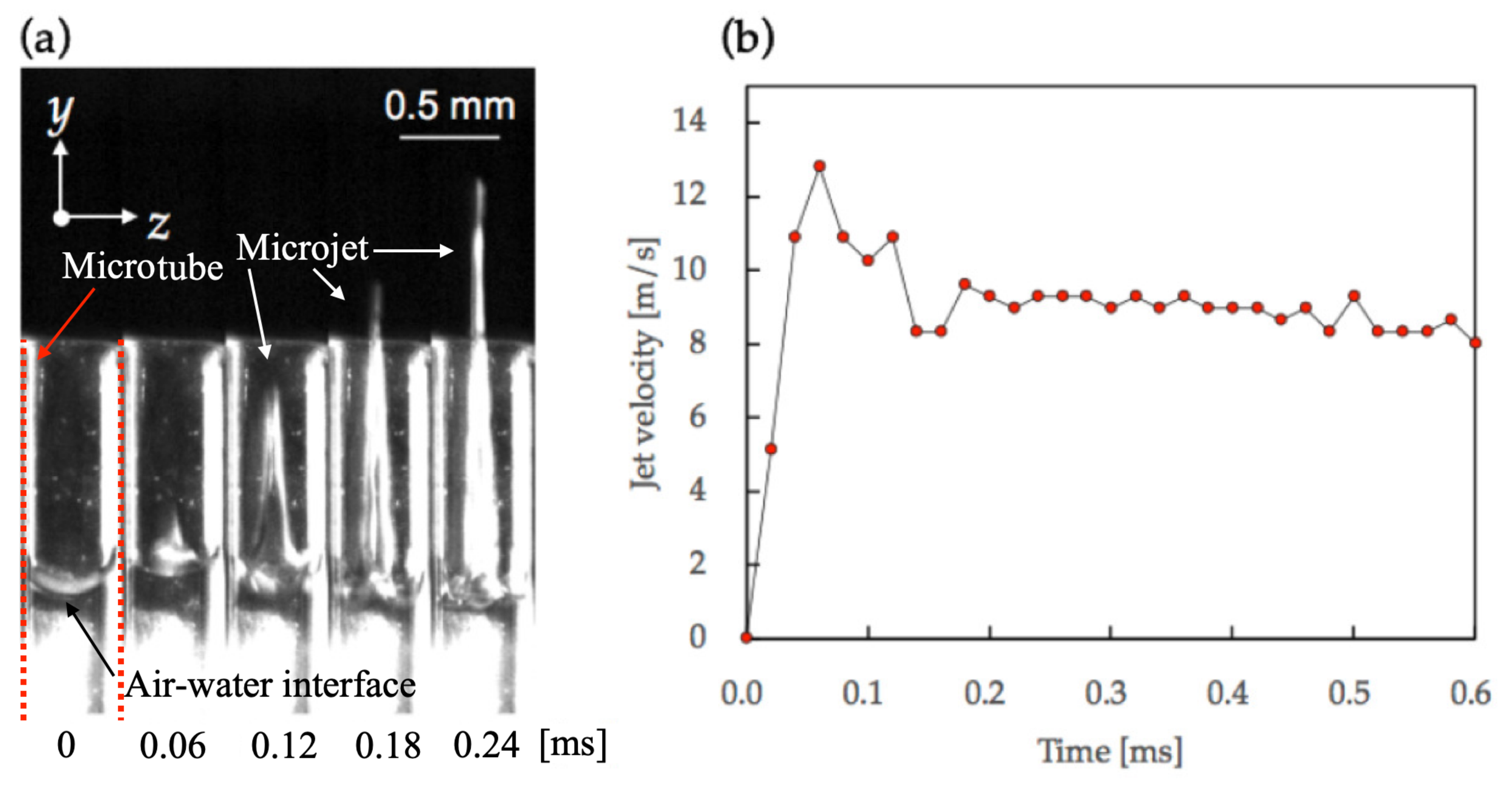}
\caption{(a) Jet generated by a laser pulse.
The subsequent images were captured at an interval of $0.06$ ms by the high-speed camera.
(b) The jet velocity was calculated from the image sequence of the jet generation shown in (a).)}
% \vspace{-5mm}
\label{fig:jet}
\end{center}
\end{figure}

Figure \ref{fig:jet}(a) presents an image sequence of the high-speed microjet generation.
The air-water interface is deformed and evolves into a focused (conical) shape.
The jet velocity as a function of time is shown in Fig. \ref{fig:jet}(b) where the jet reaches the maximum velocity of $13$ m/s when the concave air-water interface converges at $t=60$ $\upmu$s.
As previously reported by \cite{Tagawa2012}, the jet velocity decelerates due to surface tension and viscous effects, where the jet reaches an asymptotic speed of $9$ m/s.

%-----------------------------------------------------------
%-----------------------------------------------------------

\begin{figure}[!ht]
\begin{center}
\includegraphics[width=1.0\columnwidth]{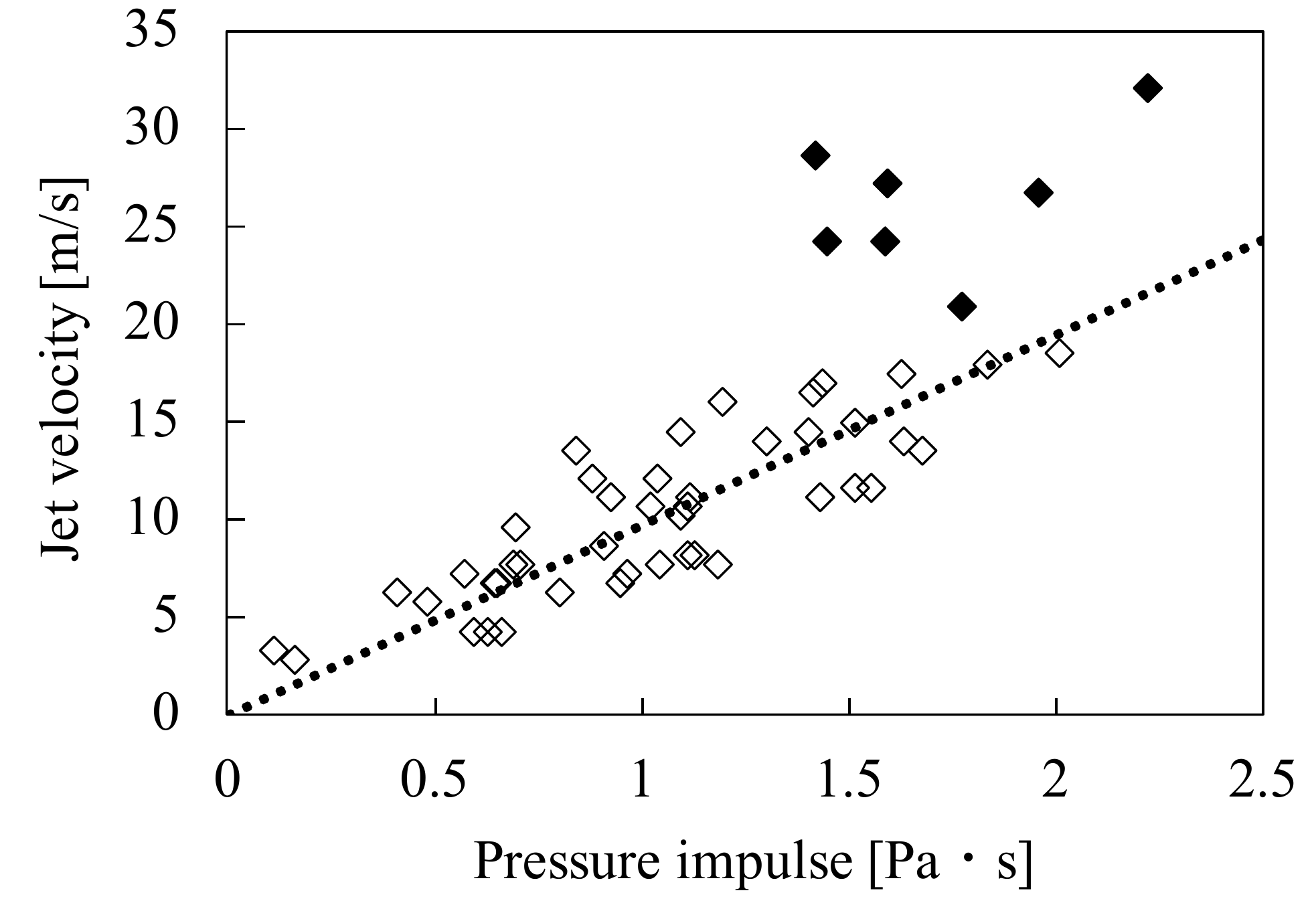}
\caption{
% Microjet speed as a function of pressure impulse, open symbols represent a system where the laser induced bubble occur and close symbols secondary cavitation is triggered.
% The fitting confirms the linear proportionality between jet speed and pressure impulse with $80$ \% of confidence as numerically predicted by Peters et al., (2013).
Microjet velocity as a function of the pressure impulse.
Opened and closed markers indicate the cases without and with secondary cavitation, respectively.
The fitted line plots the linear proportionality between the jet velocity and pressure impulse, as numerically simulated by \cite{Peters2013}, with a correlation coefficient of $0.80$.
}
% \vspace{-7mm}
\label{fig:jetspeedpressure}
\end{center}
\end{figure}

The jet velocity is plotted as a function of pressure impulse of the shock front in Fig. \ref{fig:jetspeedpressure}, where the opened and closed symbols indicate the cases without and with secondary cavitation (open markers in Fig. \ref{fig:jetspeedpressure}), respectively.
Remarkably, for the cases without secondary cavitation, the pressure impulse of the shock wave increases linearly with the jet velocity, which is consistent with theoretical and numerical studies (\cite{Peters2013}).
Thus, the high-resolution BOS technique proposed in this study has successfully measured the shock pressure in a microtube with a correlation coefficient of $0.80$ in comparison with an ideal linear model.

For the cases with secondary cavitation (closed markers in Fig. \ref{fig:jetspeedpressure}), the jet velocity is significantly higher than the fitted line, which is consistent with previous observations by \cite{Kiyama2016} and \cite{Hayasaka2017}.

\section{Conclusion}
The BOS technique is applied to an underwater shock wave generated in a micrometric rectangular capillary tube. We have visualized and quantified the shock wave by using a microdot background pattern and a high-resolution camera.
We configured our microscale BOS system with a resolution of $0.68$ $\upmu$m/pixel allowing us to measure the pressure field of a shock front of a significantly narrower width and a significantly larger peak pressure, as compared to a previous study.
The comparison of BOS pressure impulse values for the cases without cavitation and the corresponding microjet tip velocity revealed a linear relation. 
This is consistent with the theoretical predictions by \cite{Peters2013}, which has validated the accuracy of our pressure impulse measurements. 
Thus, we confirmed that our BOS technique is a suitable tool for contactless measurements of underwater shock wave pressures of microscale and potentially for pressure fields in other micro- and nanofluids by increasing the spatial-temporal resolution.
% \vspace{-2mm}

%%%%%%%%%%%%%%%%%%%%%%%%%%%%%%%%%%%%%%%%%%%%%%%%%%%%%%%%%%%%

\section*{Acknowledgement}
This work was supported by KAKENHI Grant-in-Aid for Scientific Research (A), Grant Number 20H00222/20H00223, and JST, PRESTO Grant Number JPMJPR21O5, Japan.

\bibliographystyle{ieeetr}
\bibliography{ref}
\end{document}